\documentclass[%
 reprint,
superscriptaddress,
 amsmath,amssymb,
 aps,
]{revtex4-2}

\usepackage{graphicx}
\usepackage{dcolumn}
\usepackage{bm}
\usepackage{hyperref}

\usepackage{braket}
\usepackage{subcaption}
\usepackage{xcolor}
\begin{document}

\title{Metrology of microwave fields based on trap-loss spectroscopy with cold Rydberg atoms}

\author{Romain Duverger}
\affiliation{DPHY, ONERA, Université Paris-Saclay, 91120 Palaiseau, France}

\author{Alexis Bonnin}
\affiliation{DPHY, ONERA, Université Paris-Saclay, 91120 Palaiseau, France}

\author{Romain Granier}
\affiliation{DPHY, ONERA, Université Paris-Saclay, 91120 Palaiseau, France}

\author{Quentin Marolleau}
\affiliation{DPHY, ONERA, Université Paris-Saclay, 91120 Palaiseau, France}

\author{Cédric Blanchard}
\affiliation{DPHY, ONERA, Université Paris-Saclay, 91120 Palaiseau, France}

\author{Nassim Zahzam}
\affiliation{DPHY, ONERA, Université Paris-Saclay, 91120 Palaiseau, France}

\author{Yannick Bidel}
\affiliation{DPHY, ONERA, Université Paris-Saclay, 91120 Palaiseau, France}

\author{Malo Cadoret}
\affiliation{DPHY, ONERA, Université Paris-Saclay, 91120 Palaiseau, France}
\affiliation{LCM-CNAM, 61 Rue du Landy, 93210, La Plaine Saint-Denis, France}

\author{Alexandre Bresson}
\affiliation{DPHY, ONERA, Université Paris-Saclay, 91120 Palaiseau, France}

\author{Sylvain Schwartz}
\affiliation{DPHY, ONERA, Université Paris-Saclay, 91120 Palaiseau, France}

\date{\today}

\begin{abstract}
We demonstrate a new approach for the metrology of microwave fields based on the trap-loss-spectroscopy of cold Rydberg atoms in a magneto-optical trap. Compared to state-of-the-art sensors using room-temperature vapors, cold atoms allow longer interaction times, better isolation from the environment and a reduced Doppler effect. Our approach is particularly simple as the detection relies on fluorescence measurements only. Moreover, our signal is well described by a two-level model across a broad measurement range, allowing in principle to reconstruct the amplitude and the frequency of the microwave field simultaneously without the need for an external reference field. We report on a scale factor linearity at the percent level and no noticeable drifts over two hours, paving the way for new applications of cold Rydberg atoms in metrology such as calibrating blackbody shifts in state-of-the-art optical clocks, monitoring the Earth cryosphere from space, measuring the cosmic microwave background or searching for dark matter.

\end{abstract}

\maketitle

\section{Introduction}

Rydberg atoms are a valuable resource for quantum technologies owing to their long lifetimes and large electric dipole moments between adjacent states of opposite parity \cite{adams2019rydberg, browaeys2020many}. So far, two main classes of applications have been extensively explored: quantum simulation and quantum information processing \cite{browaeys2020many}, where Rydberg interactions are harnessed to create entanglement between cold atoms controlled at the individual level, and sensing of electromagnetic fields in the radio-frequency, microwave (MW) and terahertz domains \cite{meyer2020assessment,fancher2021rydberg, wade2017real}, where the strong coupling between Rydberg atoms and external electric fields is used to perform sensitive measurements of the latter.

Typical experiments for MW field sensing with Rydberg atoms are based on the Autler-Townes splitting of an electromagnetically induced transparency (EIT) signal in a room-temperature vapor cell \cite{sedlacek2012microwave,holloway2014broadband,fan2015atom}, with state-of-the-art SI-traceable \cite{anderson2021self} measurements in the few $\mu$V/cm/$\sqrt{\textrm{Hz}}$ sensitivity range \cite{kumar2017rydberg}. Using a local MW field as a reference, it is even possible to enhance the sensitivity to a few tens of nV/cm/$\sqrt{\textrm{Hz}}$ \cite{jing2020atomic}, to measure the phase \cite{simons2019rydberg} and the angle-of-arrival \cite{robinson2021determining} of the MW field, or to perform its spectral analysis over a broad frequency range \cite{meyer2021waveguide}.

The fact that Doppler effect plays a central role in the fundamental limits of Rydberg sensors \cite{holloway2017electric,meyer2021optimal} led to a recent interest for using laser-cooled atoms in this context \cite{holloway2017electric, meyer2020assessment,liao2020microwave,zhou2023improving}. In addition to a low Doppler effect, cold atoms can be finely controlled and isolated from their environment, making them good candidates for stable and accurate measurements, as is the case for example with atomic clocks \cite{sortais2001cold, ludlow2015optical}. Moreover, the technology of cold atoms is now mature enough for field applications, as illustrated by recent demonstrations \cite{schkolnik2016compact, bidel2018absolute, antoni2022detecting, bidel2023airborne, williams2024interferometry}.

MW field measurements with cold Rydberg atoms have been reported in references \cite{liao2020microwave,zhou2023improving}, based on monitoring the transmitted intensity of a probe laser in various configurations. In particular, the work of reference \cite{liao2020microwave} demonstrated the possibility to work in a regime where the probe laser connecting the ground and the intermediate states is largely detuned from resonance, resulting in an effective two-photon coupling between the ground and the Rydberg states. In this regime, the dependence of the Autler-Townes splitting frequency versus the applied MW electric field was observed to be more linear than in the EIT regime \cite{liao2020microwave}. This is attributed to the fact that the population of the intermediate state remains negligible at all times even when the two-photon resonance condition between the lasers and the ground-to-Rydberg atomic transition is not fulfilled.

In this Letter, we report on a new method for measuring MW fields with cold Rydberg atoms, based on trap-loss spectroscopy in a magneto-optical trap (MOT). More specifically, we achieve an effective two-photon coupling between the ground state and a Rydberg state, with lasers largely detuned from the intermediate state as in reference \cite{liao2020microwave}, in a regime where the extra losses induced by blackbody ionization result in a large fluorescence change as the two-photon laser detuning is scanned across the ground-to-Rydberg atomic transition. This method is simple to implement and in particular it does not require complex detection schemes such as field ionization. Moreover, because the effective two-photon Rabi frequency of the laser coupling (on the order of 9 kHz) is much weaker than the typical MW Rabi frequency to be measured, we realize a quasi-ideal Autler-Townes configuration \cite{autler1955stark}, where the eigenstates resulting from the coupling of two Rydberg states by the MW field are only marginally perturbed by the laser fields. This simple effective two-level configuration is favorable in the context of sensing, and allows in principle the determination of the amplitude and the frequency of the MW field simultaneously from the frequencies of the two Autler-Townes spectral lines, without the need for a local oscillator. We expect this new measurement technique to be particularly well adapted to applications where accuracy, long-term stability and good resolution at large integration times are needed. The stability of our experimental setup allows to reach a spectral resolution of 20 kHz, corresponding to an equivalent field amplitude of 5 $\mu$V.cm$^{-1}$, without any observed drift over our 2 hours acquisition period. In the presence of a resonant MW field, we clearly resolve the fluctuations of our MW source.

This paper is organized as follows. We first describe the measurement principle and propose a theoretical model to account for the observed signals and the underlying physical mechanisms. We then assess some of the sensor characteristics while measuring a resonant MW signal: the linearity of the Autler-Townes splitting frequency as a function of the applied MW field amplitude, and the frequency resolution with Allan variance measurements. We then study the Autler-Townes spectra in the presence of a non-resonant MW field, showing a reasonably good agreement with the Autler-Townes formulas as long as light shifts from neighboring Rydberg states and fluctuations of our MW source are taken into account. This illustrates the possibility to measure simultaneously the frequency and the amplitude of an unknown MW field without the need of a local oscillator. We conclude by discussing future prospects of this technique in the context of electromagnetic field sensing and metrology.

\section{Measurement principle}

\subsection{Trap loss spectroscopy of $\ket{61S_{1/2}}$}

\label{sec_traploss}

\begin{figure}[h]
\begin{subfigure}[t]{0.45\textwidth}
\caption{}
\includegraphics[height = 0.55\textwidth, width=0.85\textwidth]{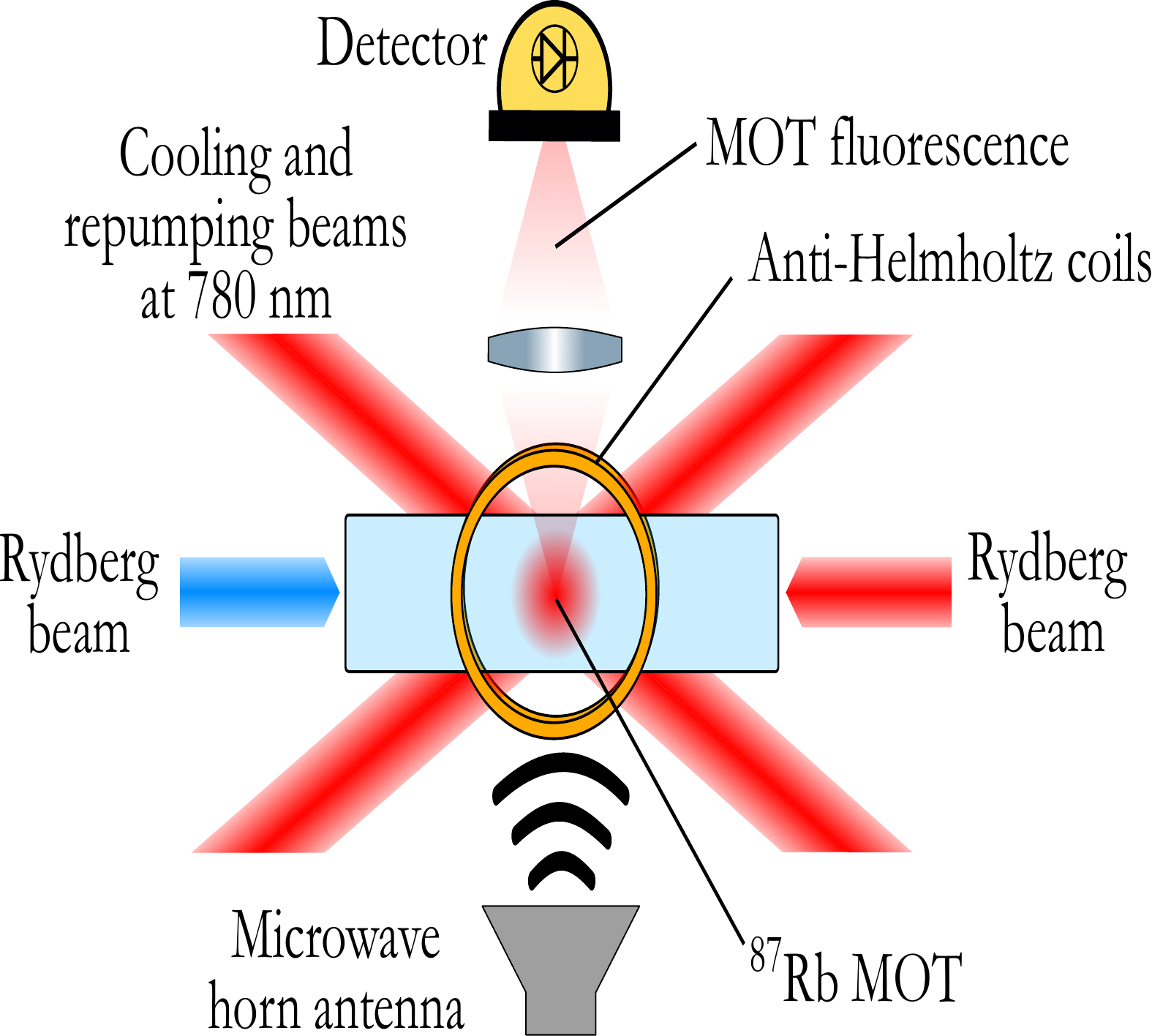}
\label{Schema manip}
\end{subfigure}
\newline
\newline
\newline
\begin{subfigure}{0.45\textwidth}
\caption{}
\includegraphics[height=0.75\textwidth, width=0.75\textwidth]{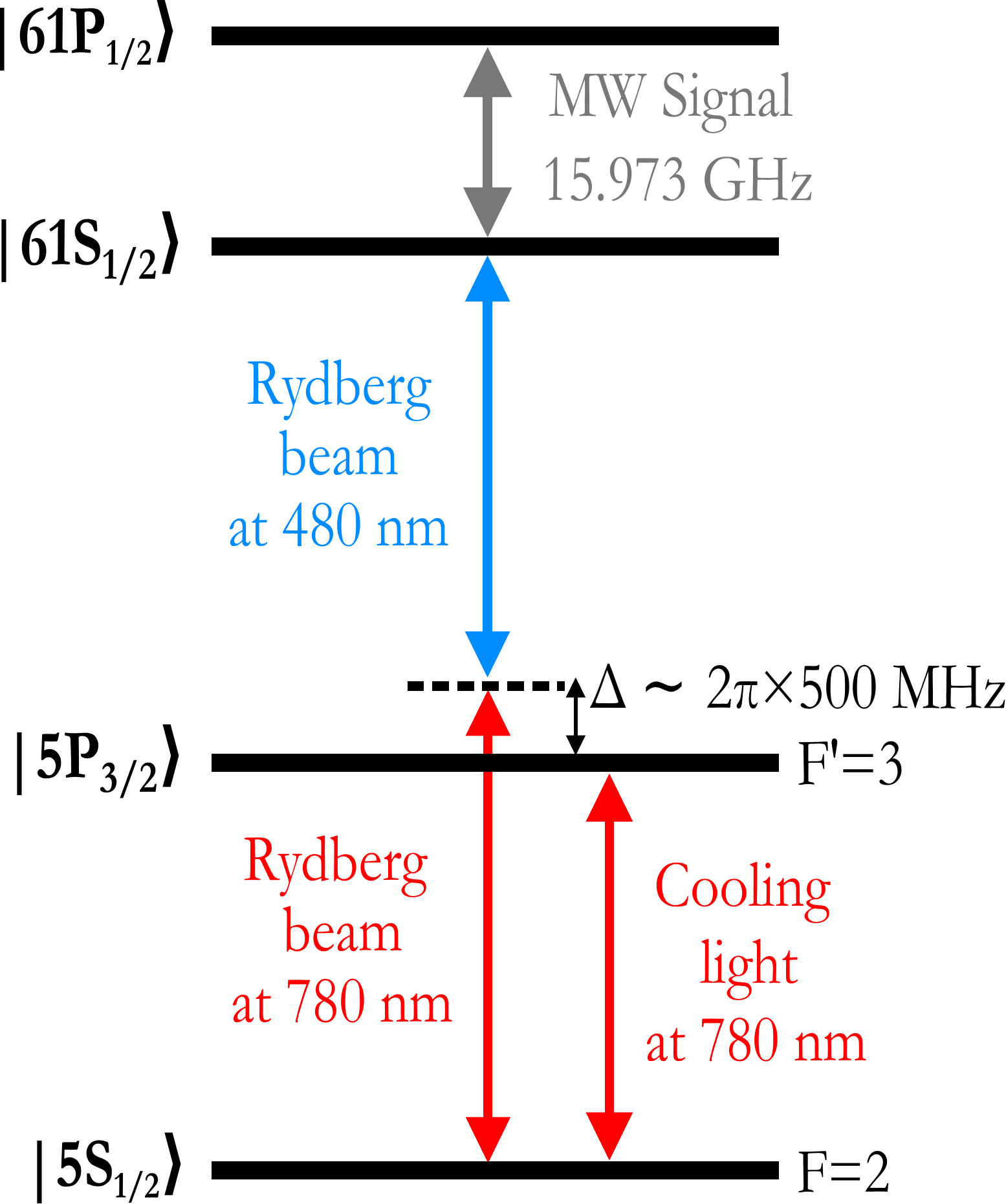}
\label{Schema niveaux}
\end{subfigure}
\caption{\label{Schemas experience} (\subref{Schema manip}) Schematic view of the experimental setup  and (\subref{Schema niveaux}) energy levels and couplings involved in the measurement process.}
\end{figure}

\begin{figure}
\centering
\begin{subfigure}{0.5\textwidth}
\caption{}
\includegraphics[height=0.64\textwidth ,  width = 0.9\textwidth]{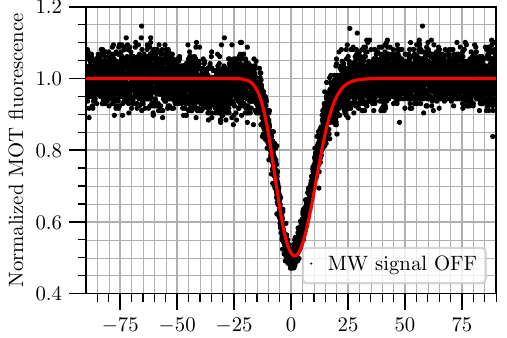}
\label{Signal trap loss MW OFF}
\end{subfigure}
\begin{subfigure}{0.5\textwidth}
\caption{}
\includegraphics[height=0.64\textwidth ,  width = 0.9\textwidth]{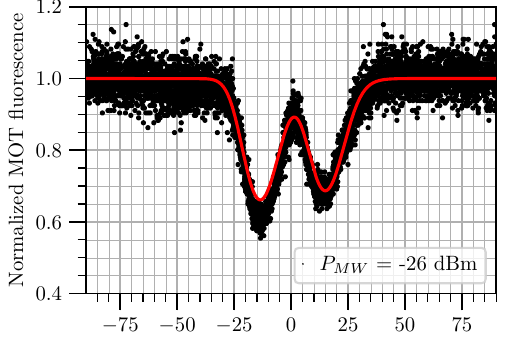}
\label{Signal trap loss MW ON 1}
\end{subfigure}
\begin{subfigure}{0.5\textwidth}
\caption{}
\includegraphics[height=0.68\textwidth, width = 0.9\textwidth]{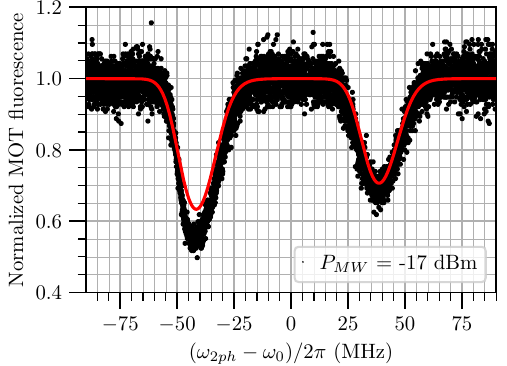}
\label{Signal trap loss MW ON 2}
\end{subfigure}
\caption{\label{Signaux trap loss} MOT fluorescence as a function of the two-photon detuning for various applied MW powers ($\omega_{2ph}$ being defined as the sum of the angular frequencies of the two Rydberg lasers). The positions of the two dips on figures \ref{Signal trap loss MW ON 1} and \ref{Signal trap loss MW ON 2} correspond to $\omega_{2ph}=\omega_\pm$ as defined by equation (\ref{AT splitting formula}). Black dots are experimental data, and red curves were derived from the rate equations model described in section \ref{SupRate}. The scanning rate is 2 MHz/s.}
\end{figure}

Our experimental setup is sketched on figure \ref{Schema manip}. We typically prepare about $10^{7}$ $^{87}$Rb atoms in a MOT. Throughout the measurement process, we leave a repumper beam on to avoid the accumulation of atoms in the hyperfine ground state $\ket{5S_{1/2},F=1}$. We couple the hyperfine state $\ket{5S_{1/2},F=2}$ to the Rydberg state $\ket{61S_{1/2}}$ using a two-photon transition with a large detuning $\Delta\sim 2\pi\times 500$ MHz from the intermediate state $\ket{5P_{3/2},F'=3}$ of natural linewidth $2\pi\times 6.065$ MHz (see figure \ref{Schema niveaux} and experimental details in \autoref{SupExp}). As we scan the two-photon detuning across the resonance between $\ket{5S_{1/2}, F=2}$ and $\ket{61S_{1/2}}$, we observe a drop in the fluorescence signal emitted by the atoms, as shown in figure \ref{Signal trap loss MW OFF}, allowing to precisely estimate the transition frequency. This technique, known as trap loss spectroscopy, has been used before in various contexts including the study of cold cesium molecules \cite{wu2011high, anderson2014photoassociation} or Rydberg series in cesium \cite{wang2007ultra, bai2019single}, strontium \cite{couturier2019measurement} and ytterbium \cite{halter2023trap}. In our case, the atom loss on resonance can be well captured by the following rate equations model \cite{day2008dynamics} taking into account losses from blackbody ionization in the Rydberg state \cite{Beterov_2009}:
\renewcommand{\arraystretch}{1.3}
\begin{equation}
\left\{
\begin{array}{ll}
\frac{dN_{g}}{dt} = L-\Gamma_{0}N_{g}- R_{exc}(N_{g}-N_{r})+\Gamma_{sp}N_{r} \\
\frac{dN_{r}}{dt} = -\Gamma_{0}N_{r}+ R_{exc}(N_{g}-N_{r})-\Gamma_\textrm{\tiny{BBR}}N_{r}-\Gamma_{sp}N_{r}
\end{array}
\right.
\end{equation}
In this model, $N_g$ and $N_r$ are the populations of the ground and Rydberg states respectively, $L$ is the MOT loading rate, $\Gamma_0 \simeq 0.6 \;\textrm{s}^{-1}$ is the overall loss rate for the atoms in the MOT in the absence of Rydberg excitation, and $R_{exc}=\Omega_{2ph}^2/(2\gamma_d)\simeq 35\;\textrm{s}^{-1}$ is the transfer rate between the ground and the Rydberg states induced by the two-photon transition with $\Omega_{2ph}\simeq 2\pi \times$ 9 kHz the effective two-photon Rabi frequency and $\gamma_d/2\pi\simeq 6.5 \times 10^6$ s$^{-1}$ the damping rate for the coherence between ground and Rydberg states, estimated from the width of the spectral profile of figure \ref{Signal trap loss MW OFF}. Finally, $\Gamma_{sp}$ and $\Gamma_\textrm{\tiny{BBR}}$ are the spontaneous emission and blackbody ionization rates, with predicted values for $\ket{61S_{1/2}}$ at room temperature on the order of $7100\;\textrm{s}^{-1}$ and $133\;\textrm{s}^{-1}$ respectively \cite{Beterov_2009}.

From these numbers, we can infer that an atom, when promoted to the Rydberg state at a rate $R_{exc}\simeq 35 \; \textrm{s}^{-1}$ on resonance, will very quickly decay back to the ground state at a rate $\Gamma_{sp}\simeq 7.1 \times 10^3 \; \textrm{s}^{-1}$, in which case it reintegrates the MOT cycle and does not contribute significantly to the loss of fluorescence. However, on some rare occasions at a rate $\Gamma_\textrm{\tiny{BBR}}\simeq 133 \; \textrm{s}^{-1}\ll \Gamma_{sp} $, an atom in the Rydberg state can be ionized by the blackbody radiation, in which case it escapes the MOT region. Because each atom spends about $R_{exc}/\Gamma_{sp}\simeq0.4\,\%$ of its time in the Rydberg state, losses from blackbody ionization will occur at a rate $\Gamma_{i}\simeq 0.004\times \Gamma_\textrm{\tiny{BBR}} \simeq 0.5 \; \textrm{s}^{-1}$. This loss rate $\Gamma_i$ comes on top of the overall loss rate for the atoms in the MOT in the absence of Rydberg excitation $\Gamma_0\simeq 0.6 \; \textrm{s}^{-1}$.
Because $\Gamma_i$ and $\Gamma_0$ have the same order of magnitude, the relative decrease of atom number and thus fluorescence $\Gamma_i/(\Gamma_0+\Gamma_i)$ can be made significantly large, $\sim 0.45$ in this case, despite the small fraction of $\simeq 0.4\,\%$ of the atoms effectively in the Rydberg states. Moreover, this leads to negligible dipolar interactions between Rydberg atoms given our atomic density of about $10^6$ mm$^{-3}$. The full width at half maximum of the spectral profile of figure \ref{Signal trap loss MW OFF} is on the order of $2.35 \times \gamma_d/2\pi \simeq 15$ MHz, which we attribute to a combination of Zeeman effect from the MOT magnetic field gradient, mixing between $\ket{5S_{1/2}, F=2}$ and $\ket{5P_{3/2},F'=3}$ by the MOT cooling light and, to a lesser extent, residual Doppler effect.

So far, we have neglected the effect of the MOT cooling light on the spectrum of figure \ref{Signal trap loss MW OFF}. It can be taken into account by a more complete model, based on optical Bloch equations (see section \ref{SupAT}). This model predicts two transitions of different strengths and frequencies between the ground and the Rydberg states, which can be interpreted as an Autler-Townes doublet resulting from the mixing of $\ket{5S_{1/2}, F=2}$ and $\ket{5P_{3/2},F'=3}$ by the MOT cooling light. As the second transition has a much lower probability, it results in a very small spectral feature (the latter is barely visible on figure \ref{Signal trap loss MW OFF}, but its presence is revealed when the numerical model is superposed to the same experimental data, as shown on figure \ref{fig6} in section \ref{SupAT}) which we neglect in the rest of this manuscript. More importantly, the MOT cooling light shifts the frequency of the main transition by about $7.6$ MHz, which we take into account in our data analysis. It should be noticed that a similar situation was studied in details in reference \cite{bai2019single}, where the coupling between ground and Rydberg states was performed by a direct single-photon excitation and where the experimental parameters led to a trap-loss spectrum with more symmetrical Autler-Townes doublet amplitudes.

Another significant effect arises from the finite response time of the MOT compared to the scanning time (more precisely the width of the spectral profile divided by the scanning rate). This leads to asymmetric spectral profiles depending on the speed of the frequency scans. To mitigate this effect, we adopt a moderate scanning rate of 2 MHz/s (unless otherwise specified) and we use asymmetric gaussian functions to fit the spectral profiles. For each measurement, we estimate the transition frequency by averaging between the two opposite scanning directions (see section \ref{SupScan} for more details).   

Throughout this manuscript, we define the origin of the frequency axis as our best estimate (see \ref{sec_stability}) of the position of the fluorescence dip while probing the $\ket{5S_{1/2},F=1} \leftrightarrow \ket{61S_{1/2}}$ transition in the absence of applied MW field, denoted $\omega_0$. In particular, $\omega_0$ includes light shifts on the ground state induced by the MOT cooling light, as discussed earlier in this section.

\subsection{Autler-Townes doublet in the presence of MW}

\label{subsection_AT_MW}

We now apply the MW signal to be measured, with a frequency at 15.973 GHz, close to resonance with the $\ket{61S_{1/2}}$ to $\ket{61P_{1/2}}$ transition \cite{vsibalic2017arc}. Typically (see figure \ref{Deltaf fonction racine puissance} and section \ref{section_linrange}), the measured MW Rabi frequency is larger than 10MHz, which is much larger than the two-photon Rabi frequency between $\ket{5S_{1/2},F=2}$ and $\ket{61S_{1/2}}$ of about $9$ kHz. Consequently, this system is very close to the textbook Autler-Townes doublet configuration, where a coupled two-level system is probed by a weak field, resulting in two spectral lines at frequencies:
\begin{equation}
\omega_\pm = \omega_0 -\frac{\Delta_{MW}}{2} \pm \frac{1}{2} \sqrt{\Omega_{MW}^{2}+\Delta_{MW}^{2}}
\label{AT splitting formula}
\end{equation}
where $\Omega_{MW}$ and $\Delta_{MW}$ are respectively the Rabi frequency and the detuning of the MW field acting on the $\ket{61S_{1/2}} \leftrightarrow\ket{61P_{1/2}}$ transition. Examples of such Autler-Townes doublets are shown on fig.\ref{Signal trap loss MW ON 1} and \ref{Signal trap loss MW ON 2}.
On resonance ($\Delta_{MW}=0$), the frequency difference between the two spectral profiles $\delta\omega = \omega_+-\omega_-$ provides a measurement of the MW Rabi frequency $\Omega_{MW}$, which is proportional to the amplitude of the MW electric field. Interestingly, the center frequency $\bar{\omega} = (\omega_++\omega_-)/2$ can be used to measure the frequency of the MW field, as will be discussed in section \ref{section_freq}. 

Even though the signals on fig.\ref{Signal trap loss MW ON 1} and \ref{Signal trap loss MW ON 2} were acquired in a quasi-resonant situation ($|\Delta_{MW}|\ll \Omega_{MW}$), the Autler-Townes doublet amplitudes are not equal. This is due to the fact that these curves were obtained by scanning the red (780nm) Rydberg laser while keeping the frequency of the blue (480nm) laser fixed, resulting in a two-photon effective Rabi frequency $\Omega_{2ph}=\Omega_1\Omega_2/2\Delta$ being higher for the line at $\omega_-$ than for the line at $\omega_+$ (see section VI.B for more details).

\section{Performance evaluation in the case of a resonant MW field}

\subsection{Linearity and measurement range} \label{section_linrange}

\begin{figure}
\centering
\includegraphics[width=0.475\textwidth , keepaspectratio=True]{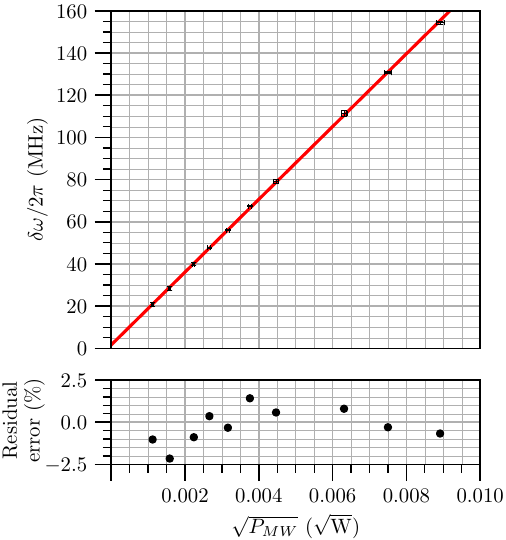}
\caption{\label{Deltaf fonction racine puissance} Measured Autler-Townes splitting as a function of the square root of the MW power at 15.973 GHz sent to the antenna. {The error bars on the x-axis correspond to an estimated uncertainty of 0.1 dB on the applied MW power. The error bars on the y-axis reflect the 2-$\sigma$ standard deviation of 10 successive measurements for each point. Their amplitude is typically smaller than 0.5~MHz.} The red curve is a linear fit to the data, yielding an intercept value of $1.6\pm 0.5$\;MHz.}
\end{figure}

The measured Autler-Townes splitting $\delta\omega$ as a function of the applied MW field is shown on figure \ref{Deltaf fonction racine puissance}. It was obtained by sending on the atoms a MW field at $15.973$ GHz, resonant with the $\ket{61S_{1/2}} \leftrightarrow\ket{61P_{1/2}}$ transition ($|\Delta_{MW}|/\Omega_{MW}\ll 1$). For each value of the power sent at the MW horn input, each data point was obtained by averaging over 10 successive pair of scans of opposite directions.  
The error bars for the $y$ axis correspond to the 2-$\sigma$ standard deviation of the 10 measurements. The error bars in $x$ correspond to an estimated 0.1 dB fluctuation on the MW power delivered to the atoms. 

As can be seen on the lower part of figure \ref{Deltaf fonction racine puissance}, the relative deviation between the experimental points and the fit is on the order of $2\,\%$ over the whole measurement range. We attribute this small discrepancy mostly to the uncertainty on the applied MW power and to the residual frequency, amplitude and polarization noise of the MOT and Rydberg lasers (see \ref{sec_stability} and \ref{SupExp}).

The intercept value resulting from the linear fit is $1.6\pm 0.5$\;MHz, corresponding to a bias in the measurement which we attribute to our fitting protocol: when the two Autler-Townes traces start to overlap, the dynamics of the MOT results in an overall trace which slightly differs from the sum of two asymetric gaussians that we use for fitting the data. This bias could be mitigated by fitting with a trace resulting from a time-dependent simulation of the rate equations as described in the supplementary sections \ref{SupRate} and \ref{SupScan}.

The minimum measurable MW splitting is set by the width of the spectral profiles shown on figure \ref{Signaux trap loss}, on the order of 15 MHz FWHM corresponding, with a theoretical dipole matrix element equal to 3164\;$ea_0$ (where $e$ is the elementary charge and $a_0$ the Bohr radius), to a MW field amplitude of about 3.8 mV/cm. As previously discussed, we attribute most of this broadening to the MOT itself: both the magnetic gradient and the mixing between $\ket{5S_{1/2},F=2}$ and $\ket{5P_{3/2},F'=3}$. 

The highest Rabi frequency $\Omega_{MW}$ that we have measured with this setup is about 160 MHz. This limitation is not fundamental, and basically comes from the fact that we scan the red Rydberg laser while keeping the blue laser fixed. Doing so, we also scan the intermediate detuning of the effective two photon transition. 
When $\Omega_{MW}$ becomes comparable to $\Delta$, the amplitude of the two Autler-Townes spectral profiles becomes too different, making it difficult to resolve the two of them precisely. To overcome this limitation, one could scan the blue Rydberg laser instead of the red, which is not currently possible on our experiment for technical reasons, or increase the detuning $\Delta$ at the expense of more laser power to maintain a similar signal to noise ratio.

\subsection{Resolution, sensitivity and long-term stability}

\label{sec_stability}

In order to evaluate the metrological performance of this measurement technique, we first record, without any applied MW signal, 57 pairs of trap-loss spectra of the ground to $\ket{61S_{1/2}}$ transition within identical experimental conditions. For each scan, the two photon laser frequency is swept at 2 MHz/s from $-50$ MHz to $+50$ MHz across the reference frequency $\omega_0/2\pi$ and back, thus setting the total measurement time to 5700 s.

\begin{figure}
\centering
\includegraphics[width=0.475\textwidth , keepaspectratio=True]{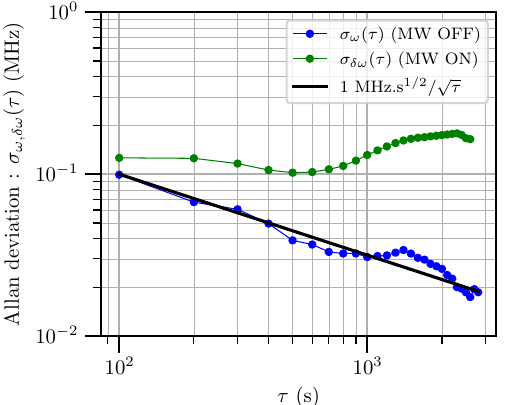}
\caption{\label{Variance d'Allan article} Overlapping Allan deviation of the transition frequency $\sigma_\omega$ in the absence of MW (blue) and of the Autler-Townes splitting frequency $\sigma_{\delta \omega}$ when a MW signal at 15.973 GHz and -23 dBm is sent to the antenna (green). The black line is a guide to the eye showing a $1/\sqrt{\tau}$ trend of the data in blue.}
\end{figure}

{The overlapping Allan deviation of the extracted transition frequency $\sigma_\omega$ as a function of the integration time $\tau$ is depicted in figure \ref{Variance d'Allan article} (blue points). We use this Allan deviation curve, taken in the absence of MW, to infer the intrinsic stability of our experiment, and extrapolate the measurement performance with the assumption that the MW induced Autler-Townes splitting can be extracted at the same level than the bare two-photon transition frequency.}

{At short-term this Allan deviation has a $1/\sqrt{\tau}$ scaling, indicating that the dominant source of fluctuation is white noise. The associated sensitivity is $\sigma_\omega \sqrt{\tau}$, which is on the order of $\simeq 1$~MHz.s$^{\frac{1}{2}}$ (or MHz/$\sqrt{\textrm{Hz}}$) in our case. This sensitivity can be converted from the frequency domain to the electric field domain using the theoretical dipole matrix element of the $\ket{61S_{1/2}}\leftrightarrow\ket{61P_{1/2}}$ transition, yielding $\simeq$ 250 $\mu$V.cm$^{-1}$.Hz$^{-1/2}$. Physically, these shot-to-shot fluctuations of the transition frequency for $\tau=100\,$s can be explained by the relative amplitude noise of 5 to $10\,\%$ on the fluorescence signals (see figure \ref{Signal trap loss MW OFF}), which mostly results from the few MHz linewidth of the cooling laser light. It should be pointed out that the measurement method described here is not optimal for maximizing the sensitivity. Indeed, a more efficient approach would be to operate the sensor only at the points where the slope of the fluorescence signal is maximal.}

{For increasing values of $\tau$, the Allan deviation curve demonstrates the capability of our setup to resolve increasingly smaller frequency changes. We define the resolution as the smallest value of the Allan deviation measured with our setup, on the order of 20 kHz for an integration time of about 2500~s as can be seen on figure \ref{Variance d'Allan article}. Once again, this frequency resolution can be converted into an electric field resolution using the appropriate transition dipole matrix element, yielding $\simeq$ 5 $\mu$V/cm.}

{We expect the long term stability of our setup to be ultimately limited by variations of the cooling light intensity, which creates lightshifts on the order of several MHz on the $\ket{5S_{1/2}}$ ground state as discussed in sections \ref{sec_traploss} and \ref{SupAT} (figure \ref{fig6}). Variations in the polarization of the cooling light and in the intensity and polarization of the Rydberg lasers might also affect the long term stability.}

We also record within the same experimental conditions 54 pairs of trap loss spectra in the presence of a MW signal. We set the estimated power and frequency sent to the antenna to $\simeq-23$ dBm and 15.973 GHz respectively, and use the Autler-Townes splitting frequency $\delta \omega/2\pi$ as the measurement signal. At short term, we obtain a slightly higher noise as the Autler-Townes doublet is slightly less contrasted than the bare spectral line (as can be seen on figure \ref{Signaux trap loss}). At longer integration times, we resolve instabilities and drifts of the MW signal sent onto the atoms. A more stable reference would be needed to fully characterize the long-term metrological performance of the current method in the presence of MW. 

\section{Experimental results in the case of a non-resonant MW field}

\label{section_freq}

\begin{figure} 
\includegraphics[width=0.45\textwidth , keepaspectratio=True]{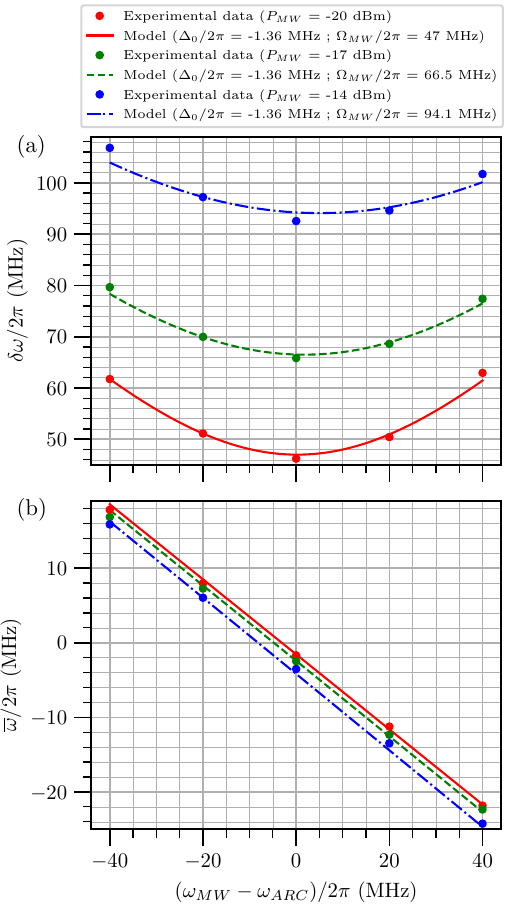}
\label{fig parabolasdroites}
\caption{\label{fig parabola full} (a) Measured Autler-Townes splitting versus MW detuning (relatively to the theoretical transition frequency of 15.973 GHz) for 3 different powers sent to the antenna (P$_{MW}$ = -20 dBm, red ; P$_{MW}$ = -17 dBm, green ; P$_{MW}$ = -14 dBm, blue). (b) Measured Autler-Townes doublet center frequency versus MW detuning from the same dataset. {In both plots, the data (color circles) are adjusted by the model of equations (\ref{eq AT with LS}) and (\ref{def de delta0}) with the following procedure: for an initial value of $\Delta_0$ we perform three independent adjustments for the datasets associated to -20 dBm, -17 dBm and -14 dBm respectively to find three values of $\Omega_{MW}$. We then iterate on the value of $\Delta_0$ and repeat the procedure until the distance between the fits and the datasets is minimized. This procedure yields to the following values: $\Delta_0=-2\pi \times 1.36\,$MHz and $\Omega_{MW} = 2\pi \times 47\,$MHz (red dashed lines, P$_{MW}$ = -20 dBm); $2\pi \times 66.5\,$MHz (green dashed lines, P$_{MW}$ = -17 dBm); $2\pi \times 94.1\,$MHz (blue dashed line, P$_{MW}$ = -14 dBm).} The fact that the three lines on figure (b) have different offsets is a consequence of light shifts, see appendix \ref{SupStark} for more details.}
\end{figure}

So far, we have considered only the situation where the MW field resonantly couples the two Rydberg states $\ket{61S_{1/2}}$ and $\ket{61P_{1/2}}$, resulting in a quasi-linear dependence of the Autler-Townes frequency versus the MW field amplitude. In this section, we now consider the situation where the MW field can be detuned from the $\ket{61S_{1/2}}\leftrightarrow\ket{61P_{1/2}}$ transition, resulting in a non-linear dependence as predicted by equations (\ref{AT splitting formula}). In this context, an extra ingredient that we have neglected so far must be included in the model, namely the existence of light shifts on the $\ket{61S_{1/2}}$ and $\ket{61P_{1/2}}$ states induced by MW couplings to neighboring Rydberg states. The most important shift results from the $\ket{61S_{1/2}}\leftrightarrow\ket{61P_{3/2}}$ coupling as it is only 437 MHz detuned from the targeted MW transition and exhibits a higher dipole matrix element of 4404\;$ea_0$. We also take into account in our model the shifts induced by the couplings between $\ket{61S_{1/2}}$ and \{$\ket{60P_{1/2}}$ ; $\ket{60P_{3/2}}$\} and between $\ket{61P_{1/2}}$ and \{$\ket{62S_{1/2}}$ ; $\ket{60D_{3/2}}$ ; $\ket{59D_{3/2}}$\}. In this case, the Autler-Townes formula (\ref{AT splitting formula}) can be rewritten in terms of the sum and difference of the frequencies of the Autler-Townes spectral lines as:
\begin{equation}
\left\{
	\begin{array}{ll}
		\delta\omega = \omega_+ - \omega_- = \sqrt{\Omega_{MW}^2 + (\Delta_{MW} + \delta_1^{LS} - \delta_2^{LS})^2}\\
		\bar{\omega} = \frac{1}{2}(\omega_+ + \omega_-)  - \omega_0 = -\frac{1}{2}(\Delta_{MW} - \delta_1^{LS} - \delta_2^{LS})
	\end{array}
\right.
\label{eq AT with LS}
\end{equation}
where $\delta_{1,2}^{LS}$ are the overall light shifts on the $\ket{61S_{1/2}}$ and $\ket{61P_{1/2}}$ states respectively. The theoretical calculation and experimental measurement of $\delta_{1,2}^{LS}$ is described in appendix \ref{SupStark}. To give a sense of scale, for a MW power corresponding to $\Omega_{MW} = 2\pi\times 100\;\textrm{MHz}$, we find $\delta_{1}^{LS} \simeq 2\pi\times(-7.4)\;\textrm{MHz}$ and $\delta_{2}^{LS} \simeq 2\pi\times(-0.4)\; \textrm{MHz}$ respectively.

We report on figure \ref{fig parabola full} the experimentally measured values of $\delta\omega$ and $\bar{\omega}$ for various applied values of $\Omega_{MW}$ and $\Delta_{MW}$, showing a reasonably good agreement with the model from equations (\ref{eq AT with LS}). In order to account for a possible uncertainty on the bare $\ket{61S_{1/2}}\leftrightarrow\ket{61P_{1/2}}$ transition frequency, we write the MW detuning in the absence of light shifts as:
\begin{equation} \label{def de delta0}
\Delta_{MW}=\omega_{MW}-\omega_{ARC}-\Delta_0 \; ,
\end{equation}
where $\omega_{MW}$ is the applied MW frequency, $\omega_{ARC}=2\pi \times 15.973$\;GHz is the atomic transition frequency from ARC Rydberg calculator \cite{vsibalic2017arc} and $\Delta_0\simeq -2\pi \times 1.36$\;MHz according to the adjustment of the data by the model as shown on figure \ref{fig parabola full}. We attribute the origin of the residual discrepancy between the data and the model (up to a few MHz at most) to be mainly due to fluctuations in the applied MW power. This is supported by the fact that the $\delta\omega$ curves (figure \ref{fig parabola full}), which have a much stronger dependence in the MW power than $\bar{\omega}$ according to the theory, also have a much higher level of discrepancy. Other possible causes for the observed discrepancy include: the deterioration of the signal to noise ratio for one of the two spectral profiles as the value of the detuning increases and becomes comparable with $\Omega_{MW}$; the bias induced by our fitting protocol as discussed in section \ref{section_linrange}; the effects of the finite response time of the MOT (section \ref{SupScan}), which might be enhanced when the amplitudes of the two spectral profiles become significantly different; the uncertainties in the estimation of the dipole matrix elements for the light shift calculation, which might become more and more important for increasing values of the MW power.

Noticeably, because the lighshifts are much smaller than $\Omega_{MW}$, there is a one-to-one correspondence between a couple $(\delta\omega,\bar{\omega})$ and a couple $(\Omega_{MW},\Delta_{MW})$, which means that from a single Autler-Townes measurement one can simultaneously deduce the amplitude and the frequency of the MW field, without the need for a local oscillator \cite{meyer2021waveguide}.

\section{Conclusion}

In this paper, we have demonstrated and characterized a new method to measure MW fields with Rydberg atoms based on trap-loss spectroscopy in a magneto-optical trap. This method is particularly simple as the detection scheme relies on fluorescence measurements only. By using a two-photon transition highly-detuned from the intermediate state, we realize a situation where the frequencies of the spectral lines are well-described by a coupled two-level system, which is particularly favorable for the linearity of the sensor in the resonant case. In the non-resonant case, this simple two-level system behavior allows in principle to reconstruct both the amplitude and the frequency of the applied MW field from the Autler-Townes splitting frequency and doublet center frequency, provided the light shifts induced by the MW field are taken into account. {The maximum range that we achieved for the measurement of the microwave power, on the order of 160~MHz,} is already much higher than for electromagnetically-induced absorption in cold atoms \cite{liao2020microwave}, and could be possibly much larger if the coupling (blue) laser was scanned instead of the probe (infrared) laser, which could be technically facilitated by the use of the inverted scheme \cite{urvoy2013optical} where the coupling laser is in the infrared domain (1010nm). {The maximum range that we achieved for the measurement of the microwave frequency is on the order of $\pm 50$~MHz around the resonance frequency, typically set by the value of the MW Rabi frequency on resonance.} Future directions to improve the metrological performance of our experimental setup include a better control of the intensity of the cooling beams, Rydberg beams and their associated polarizations, as well as a better control of the MW power effectively sent onto the atoms. In the longer term, one could get rid of the MOT broadening by adapting the setup to perform pulsed MOT operation, optical molasses or even dipole traps.

{In comparison with state-of-the-art techniques based on Rydberg EIT in room-temperature vapor cells, this new approach has several advantages, including reduced Doppler and transit time effects, better long-term stability as the atoms are more isolated from their environment, and the possibility to measure simultaneously the amplitude and the frequency of the electric field. On the other hand, the time needed to perform a single measurement with this new approach is much higher than for vapor cells as it is limited by the response time of the magneto-optical trap, on the order of one second in the present work.}

With a long-term frequency stability equivalent to a resolution of 5$\mu$V/cm at 2500s and no noticeable drift over this time period, this new measurement technique appears to be particularly well-suited for metrology experiments where accuracy, long term stability and high resolution at large integration times are required. This includes in particular technological applications such as the characterization of radiofrequency and MW equipment or new concepts of radar or RF antennas where the long term stability would be brought by cold atoms and the high bandwidth by Rydberg EIT in a hot vapor cell. Controlling the external degrees of freedom of the cold atoms could also be used for high-resolution THz imaging, or to extend the measurement range of Rydberg sensors by making different atoms resonant with different microwave frequencies, for example using a gradient of electric field. This platform also holds great prospects for scientific applications such as black-body shifts measurements in state-of-the-art optical clocks \cite{ovsiannikov2011rydberg}, monitoring the Earth cryosphere from space \cite{strangfeld2023quantum}, measuring the cosmic MW background \cite{tscherbul2014coherent} or searching for axions or other forms of dark matter \cite{shibata2008practical, gue2023search}.

\section{Appendix}

\subsection {Experimental setup details}
\label{SupExp} 

We create a $^{87}$Rb MOT in a glass cell with a 26 G/cm magnetic field gradient and 3 pairs of retro-reflected beams at 780 nm with 1/e$^{2}$ diameter around 9 mm, used for both cooling and repumping. {We estimate the partial pressure of rubidium in the cell to be on the order of $10^{-8}$ Torr}. The dimensions of the atomic cloud are typically on the order of 1 mm. The cooling beams come from a single Ortel telecom laser diode at 1560 nm, frequency-doubled with a PPLN waveguide crystal. The frequency of the cooling light is detuned by 20 MHz from the $\ket{5S_{1/2},F=2} \leftrightarrow \ket{5P_{3/2},F'=3}$ transition. The repumping light comes from the phase modulation of the cooling light, with a power ratio beetween the repumping and cooling frequencies around 20 \%. The MOT beams carry a total amount of power around 180 mW. The coupling between the ground state and the $\ket{61S_{1/2}}$ Rydberg state is performed with a 2 photon scheme (780 nm and 480 nm), with about 500 MHz detuning from the intermediate $\ket{5P_{3/2},F'=3}$ state. Both beams are linearly polarized, their direction of polarization being adjusted to maximize the contrast of the spectral feature shown in figure \ref{Signal trap loss MW OFF}. The two counterpropagating Rydberg laser beams have a $1/e^{2}$ diameter of 1 mm, and their respective power are 2 mW for the infrared one (frequency-doubled Rio Planex telecom laser diode at 1560 nm) and 80 mW for the blue one (Moglabs injection-locked ILA system). The two lasers are frequency-stabilized using a Pound-Drever-Hall lock on an ultrastable cavity from SLS with a finesse of the order of 2100 for the infrared light and 18000 for the blue one. The sweep of the two-photon detuning is achieved by locking a sideband of the phase-modulated infrared Rybderg laser to the reference cavity while sending the carrier to the atoms, and then sweeping the frequency of the phase modulation. The MW signal sent onto the atoms is provided by a Rohde \& Schwarz SMB100A generator stabilized by a 10 MHz reference from a GPS receiver. The horn antenna is set 10 cm away from the atoms.

\subsection {Theoretical model and rate equations}
\label{SupRate}

In the context of a two-photon transition, we consider only the two-level system composed of states \(\ket{g} \equiv \ket{5S_{1/2},F=2}\) and \(\ket{r} \equiv \ket{61S_{1/2}}\). The theoretical value from the literature of the state $\ket{r}$ lifetime is \(\tau = 140\,\mu\)s \cite{vsibalic2017arc}, leading to a spontaneous emission rate \(\Gamma_{sp}=\frac{1}{\tau} \sim 7.1\times10^{3}\;\textrm{s}^{-1}\). The damping rate for the dipole coherence \(\gamma_{d}\) is estimated from the experimentally measured width of the spectral feature of figure \ref{Signal trap loss MW OFF} to be \(\gamma_{d}/2\pi \sim 6.5 \times 10^{6}\,s^{-1}\). This is consistent with the back-of-the-envelope calculation of the Zeeman spectral width across the 1-mm MOT with a magnetic field gradient of {\(\nabla B \sim 26\,\textrm{G/cm}\)}, yielding to a few MHz. In this regime of strong damping of dipole relaxation (\(\gamma_{d} \gg \Gamma_{sp}\)), the dipole adiabatically follows the temporal evolution of populations, enabling the elimination of the coherence term in the optical Bloch equations and simplifying the model to two coupled rate equations. We furthermore consider a MOT loading rate \(L\), that affects the population ground state \(N_{g}\), and loss processes \(\Gamma_{0}\) affecting both states population as \(-\Gamma_{0}N_{g}\) and \(-\Gamma_{0}N_{r}\). 
Atoms in the Rydberg state may exit the trap due to processes such as photo-ionization or blackbody ionization, which is modeled as \(-\Gamma_\textrm{\tiny BBR}N_{r}\). According to reference \cite{Beterov_2009}, a theoretical estimate of \(\Gamma_\textrm{\tiny BBR}\) is \(\simeq 133\;\textrm{s}^{-1}\). The terms for absorption and stimulated emission are given by \(\pm R_{exc}(N_{g}-N_{r})\), with $R_{exc}=R_0=\Omega_{2ph}^{2}/2\gamma_d$ on resonance. When the two-photon detuning $\delta$ is non zero, we write the excitation rate as \(R_{exc}(\delta)=\frac{R_{0}}{(1+\frac{\delta}{\Delta})^{2}} \times \mathcal{P}(\delta)\), in order to take into account the explicit dependency \(R_{exc} \propto \Omega_{2ph}^{2} \propto \frac{1}{(\Delta+\delta)^{2}}\) and an excitation profile $\mathcal{P}(\delta)$ characterizing the shape of the resonance. Even though the adiabatic approximation of the optical Bloch equations (for an isolated single atom) theoretically yields a Lorentzian excitation profile $\mathcal{P}(\delta)$, our experimental data are much better described by a Gaussian model \(R_{exc}(\delta)=\frac{R_{0}}{(1+\frac{\delta}{\Delta})^{2}}\times e^{-\frac{\delta^{2}}{2\gamma_{d}^{2}}}\), which accounts for the inhomogeneous broadening across the MOT. We finally obtain :
\begin{equation} \label{rate_eq}
\left\{
\begin{array}{ll}
\frac{d}{dt}N_{g} = L-\Gamma_{0}N_{g}- R_{exc}(\delta)(N_{g}-N_{r})+\Gamma_{sp}N_{r} \\
\frac{d}{dt}N_{r} = -\Gamma_{0}N_{r}+ R_{exc}(\delta)(N_{g}-N_{r})-\Gamma_\textrm{\tiny BBR}N_{r}-\Gamma_{sp}N_{r}
\end{array}
\right.
\end{equation}

This model was used to calculate the red traces on figure \ref{Signaux trap loss} (additionally taking into account, in the presence of MW, the relative strengths of the spectral lines from the Autler-Townes theory \cite{autler1955stark}), leaving \(R_{0}\) as a free parameter which best fits the experimental data for \(R_{0}=35\,\textrm{s}^{-1}\). This enables us to derive an estimation for the average two-photon Rabi frequency experienced by the atoms, using $R_{0} = \Omega_{2ph}^{2}/2\gamma_d$ (note that this expression naturally arises in the two-level population model when writing the Bloch equations under the adiabatic approximation). We find \(\Omega_{2ph} \sim 2\pi \times 9\;\textrm{kHz}\), which is consistent (given the fact that the atomic cloud is slightly bigger that the beam waists) with the theoretical value calculated at the point of maximum intensity of the gaussian Rydberg laser beams, the latter yielding \(\Omega_{780nm}=2\pi \times 41\;\textrm{MHz}\), \(\Omega_{480nm}=2\pi \times 0.57\;\textrm{MHz}\) and finally \(\Omega_{2ph}=\Omega_{780nm}\Omega_{480nm}/2\Delta=2\pi \times 23\;\textrm{kHz}\) for $\Delta=2\pi \times 500$ MHz.

\subsection {Autler Townes effect caused by the cooling light}
\label{SupAT}

\begin{figure}
\centering
\includegraphics[width=0.475\textwidth, keepaspectratio=True]{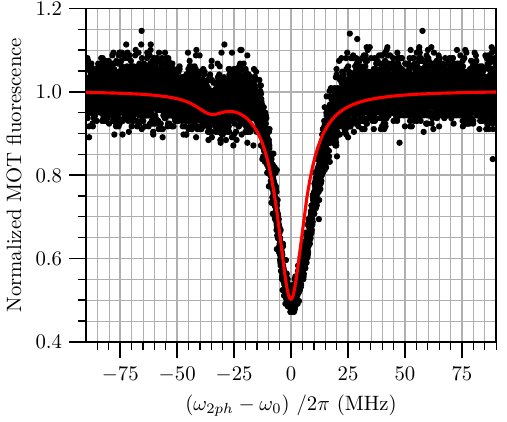}
\caption{\label{fig6} Autler Townes splitting due to the coupling of $\ket{5S_{1/2},F=2}$ and $\ket{5P_{3/2},F'=3}$ by the cooling light. The red trace is the theoretical model and the black dots are the experimental data already presented on figure \ref{Signal trap loss MW OFF}. The origin of frequencies for the red trace has been shifted by about $7.6$ MHz to match the experimental data, corresponding to the shift induced by the MOT cooling light on the trap loss spectral features.}
\end{figure}

As discussed in section \ref{sec_traploss}, a close look at the experimental curve of figure \ref{Signal trap loss MW OFF} reveals the presence of a second spectral feature around $-35$ MHz. This feature can be explained by the cooling light operating on the \(\ket{5S_{1/2}, F=2} \leftrightarrow \ket{5P_{3/2}, F'=3}\) transition with a Rabi frequency \(\Omega_{MOT}=2\pi \times 29\,\textrm{MHz}\) and a detuning \(\delta_{MOT}=2\pi \times -20\,\textrm{MHz}\), which results in an Autler-Townes splitting of the ground state also involved in the trap loss measurement. This is quantitatively confirmed by a numerical model based on Bloch equations and involving the three atomic states \(\ket{g} \equiv \ket{5S_{1/2},F=2}\), \(\ket{i} \equiv \ket{5P_{3/2},F'=3}\) and \(\ket{r} \equiv \ket{61S_{1/2}}\), and two Rabi couplings: \(\Omega_{MOT}\) between \(\ket{g}\) and \(\ket{i}\), and \(\Omega_{2ph}=2\pi \times 9\;\textrm{kHz}\) between \(\ket{g}\) and \(\ket{r}\). By furthermore taking into account the decay rates \(\Gamma_{\ket{i}}=2\pi \times 6.06 \times 10^{6}\,s^{-1}\) and \(\Gamma_{\ket{r}}=7.1 \times 10^{3}\,s^{-1}\), and by introducing empirical coherence damping rates \(\gamma_{gi}=2\pi \times 2\) MHz and \(\gamma_{gr}=2\pi \times 4\) MHz, we obtain the red trace on figure \ref{fig6} which shows a good agreement with the measured data (identical to the one presented on figure \ref{Signal trap loss MW OFF}).

Interestingly, in a similar work reported in reference \cite{bai2019single}, the two Autler-Townes features from the MOT cooling light had comparable sizes, which we attribute to different experimental parameters (Rabi frequency and detuning of the cooling light, and a much higher Rabi frequency for the ground-to-Rydberg transition owing to direct single-photon coupling).

\subsection{Influence of the MOT dynamics}

\label{SupScan}

\begin{figure}
\centering
\includegraphics[width=0.48\textwidth, keepaspectratio=True]
{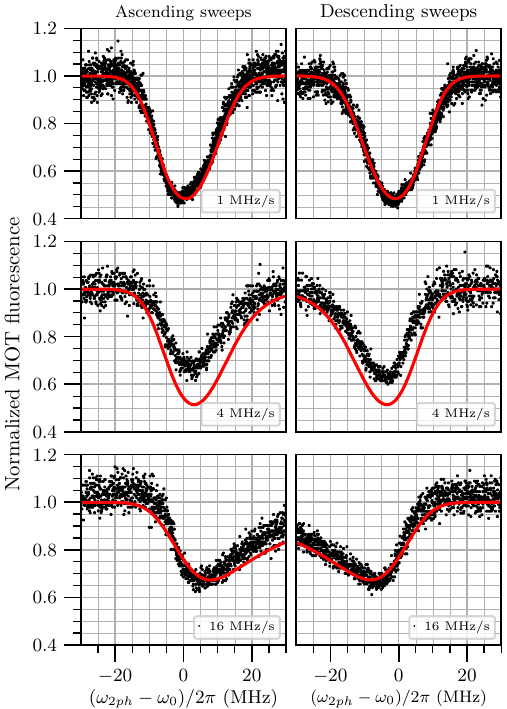}
\caption{\label{Sweep rate trap loss} MOT fluorescence as a function of the two-photon detuning for various scanning rates (1 MHz/s, 4 MHz/s and 16 MHz/s from top to bottom). The first (second) column corresponds to sweeps of increasing (decreasing) frequencies. The back dots are experimental data, and the red traces result from a rate equation model (see text).}
\end{figure}

As we scan the frequency of the two-photon coupling across the resonance between the ground and the Rydberg states, the fluorescence of the MOT changes at a rate $\Gamma_0 \simeq 0.6 \; \textrm{s}^{-1}$. Because of the finite response time $\Gamma_0^{-1}$, the shape of the fluorescence trace will depend on the rate at which the scan is performed. This is illustrated on figure \ref{Sweep rate trap loss}, where it can be seen that the traces become more an more asymmetric as the scanning rate is increased. This asymmetry, which also depends on the direction of the scan, is reasonably well reproduced by the numerical integration of the rate equations (\ref{rate_eq}) with a time-dependent detuning $\delta$ (red curves on figure \ref{Sweep rate trap loss}). The frequency corresponding to the minimum value of the fluorescence can be estimated by fitting each trace with a skewed gaussian function (or a double skewed gaussian function in the case of an Autler-Townes doublet). 

\begin{figure}
\centering
\includegraphics[width=0.48\textwidth, keepaspectratio=True]
{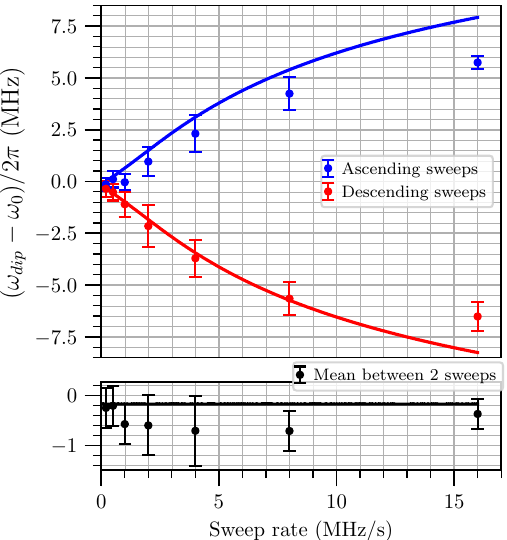}
\caption{\label{Deltaf fonction sweep rate} Estimated frequency corresponding to the minimum of each fluorescence trace for various scanning rates and directions. The dots are experimental data, and the curves theoretical predictions from the rate equation model (see text).}
\end{figure}

We report on figure \ref{Deltaf fonction sweep rate} the estimated frequency $\omega_{dip}/2\pi$ corresponding to the minimum fluorescence for each trace, as a function of the scanning rate and direction. As expected the estimation error increases with the scanning rate, but the average between the estimated frequencies using an ascending and a descending scan remains stable at the MHz level. The estimated frequencies based on the theoretical traces, also shown on figure \ref{Deltaf fonction sweep rate}, are in good agreement with the experimental points. Interestingly, the value of $\omega_{dip}/2\pi$ predicted by the model in the limit of very slow scanning rates is not zero but $\simeq 200$ kHz. This is due to the fact that the Rabi frequency $\Omega_{2ph}$ of the effective two-photon coupling between the ground and Rydberg states has a dependence in $\delta$ (see section \ref{SupRate}), breaking the symmetry between ascending and descending scans.

Throughout the manuscript, we used a scanning rate of 2 MHz/s as a trade-off between the asymmetry and contrast of the fluorescence trace and the data acquisition time. Each presented measurement is the result of an average over one ascending and one descending scan.

\subsection {Additional detuning from the MW source}

\label{SupStark}

The MW signal at 15.973 GHz which resonantly couples $\ket{61S_{1/2}}$ and $\ket{61P_{1/2}}$ also results in an energy shift of these two states due to off-resonant coupling to other neighboring Rydberg states. To evaluate this effect, we use the following formula \cite{grimm2000optical}:
\begin{equation}
\delta_{i}^{LS} = \sum_j \frac{|\Omega_{ij}|^2}{4\Delta_{ij}}
\label{Light shift expression}
\end{equation}
where $i=1,2$ refers to the $\ket{61S_{1/2}}$ and $\ket{61P_{1/2}}$ respectively, and $j$ refers to all other states coupled to 1 and 2 by the MW field. In the above formula, $\Omega_{ij}$ is the Rabi frequency for the $i\leftrightarrow j$ transition, and $\Delta_{ij}=\omega_{MW}-\omega_{ij}$ the detuning between the MW field and the $i\leftrightarrow j$ transition. To estimate $\Omega_{ij}$, we use the theoretical value of the transition dipole element $d_{ij}$ between $i$ and $j$ (\emph{Reduced Matrix Element J} from ARC calculator \cite{vsibalic2017arc}), and the following formula: $\Omega_{ij}=d_{ij}\Omega_{12}/d_{12}$ (with previous notations, $\Omega_{12}=\Omega_{MW}$). We keep only the states $j$ whose contribution to either $\delta_{1}^{LS}$ or $\delta_{2}^{LS}$ is significant (typically $>100$ kHz for the reference value $\Omega_{12}=2\pi \times 100$ MHz). Such states are 
$\ket{61P_{3/2}}$, $\ket{60P_{3/2}}$ and $\ket{60P_{1/2}}$ for $\delta_{1}^{LS}$, and $\ket{62S_{1/2}}$, $\ket{60D_{3/2}}$ and $\ket{59D_{3/2}}$ for $\delta_{2}^{LS}$.

\begin{figure}
\centering
\includegraphics[width=0.48\textwidth , keepaspectratio=True]{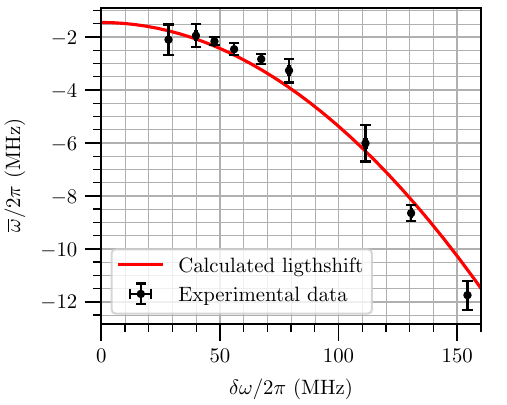}
\caption{\label{Pos fonction racine puissance} Average position of the Autler-Townes doublet versus MW power. The red line is the calculated light shift $(\delta_{1}^{LS}+\delta_{2}^{LS})/2$ with a $-1.5$ MHz offset to account for the fact that the MW frequency is not perfectly resonant with the bare $\ket{61S_{1/2}} \leftrightarrow \ket{61P_{1/2}}$ transition.}
\end{figure}

We show on figure \ref{Pos fonction racine puissance} the calculated light shift $(\delta_{1}^{LS}+\delta_{2}^{LS})/2$ as a function of the MW Rabi frequency, showing a good correlation with the average position of the Autler-Townes doublet as expected from equations (\ref{eq AT with LS}).


\section*{Acknowledgement}

We acknowledge funding from Agence Nationale de la Recherche within the project ANR-22-CE47-0009-03 and from Investissements d’Avenir du LabEx PALM within the project ANR-10-LABX-0039-PALM. We thank Nathan Bonvalet, Nicolas Guénaux and Guillaume de Rochefort for early contributions to the experimental work and simulations, and the electromagnetism and radar department (DEMR) of ONERA for fruitful discussions and for lending us microwave equipment. Romain Granier acknowledges funding from Ecole Normale Supérieure Paris-Saclay through a CDSN doctoral grant.

\end{document}